# Evaluation of biometric user authentication using an ensemble classifier with face and voice recognition

**Firas Abbaas and Gursel Serpen**

Department of Electrical Engineering & Computer Science, University of Toledo,
Toledo, Ohio 43606, USA
gursel.serpen@utoledo.edu

*Abstract*: **This paper presents a biometric user authentication system based on an ensemble design that employs face and voice recognition classifiers. The design approach entails development and performance evaluation of individual classifiers for face and voice recognition and subsequent integration of the two within an ensemble framework. Performance evaluation employed three benchmark datasets, which are NIST Feret face, Yale Extended face, and ELSDSR voice. Performance evaluation of the ensemble design on the three benchmark datasets indicates that the bimodal authentication system offers significant improvements for accuracy, precision, true negative rate, and true positive rate metrics at or above 99% while generating minimal false positive and negative rates of less than 1%.**

## I. Introduction

A majority of secure systems still employ a user authentication procedure as the method of determining the identity of an individual. This process entails a text-based password entry: such authorization methods are appealing from the viewpoint of computational complexity. They result in minimal computational cost in terms of space and time as it only entails capturing the password from the user and comparing it with a pre-stored word or phrase in a database. The major disadvantage of the password authorization method is the low-level reliability and potential vulnerability, particularly for so-called "weak" passwords associated with determining the actual identity of the user. In other words, password authorization grants access to any user who happens to "know" the correct password. Therefore, in systems for which security is a critical aspect, and the actual identity of the user needs to be established with high levels of confidence, biometric authentication approach offers an option.

Biometric authentication is the method of determining the identity of an individual based on the inherent physical or behavioral traits associated with that person [20,21,27-29, 30-36]. It leverages a variety of methods that may utilize fingerprints, iris, face, hand geometry, voice, signature, keyboard typing pattern, etc. to be able to recognize an individual. It provides the strongest link between the actual user and the system. Generally, a biometric authentication system functions by capturing the biometric trait of a person and comparing the recorded trait with the biometric samples of the same person which are previously captured in a database in order to establish the identity of that individual. The need for establishing identity in a reliable manner for highly secure systems has spurred active research in the field of biometrics [22]. Unlike the traditional password authentication, biometric authentication processes (such as face and voice recognition) require significantly more computation power than what is needed for password authentication.

Biometric systems may be classified into two types; unibiometric and multibiometric authentication systems. A unibiometric system is one that depends on a single biometric source (such as voice or face biometric traits) for user authentication. The multi biometric system depends on multiple biometric sources fusing them into a single authentication decision. In general, biometric data are vulnerable to distortion or corruption due to environmental factors. For instance, significant lighting variations can make the face of an individual "look" completely different to the authenticating device. Therefore, a unibiometric system is usually not a good solution, as it can be highly susceptible to performance degradation due to environmental conditions. Another relevant point is that individual biometric traits are each affected by typically different environmental conditions: face by light, voice by noise, fingerprint by skin conditions etc. Therefore, a multi biometric authentication system offers a promising option as it considers more than one biometric trait, potentially increasing reliability of the authentication process.

This study presents a multimode biometric authentication system design, development and performance evaluation through simulation. The proposed system is bimodal employing two unimode biometric authentication systems, one based on face recognition and a second one based on voice recognition. The design approach is such that these two unimode systems are initially developed individually as machine learning classifiers. Next, a fusion module is developed to form a multimode system as an ensemble classifier to combine or fuse these two unibiometric module output or decisions into a single one with high confidence.
In the forthcoming sections, development of the face identification system is presented first followed by the development of the voice identification system. The design of a fusion system along with performance assessment and evaluation of face identification, voice identification and the overall bimodal biometric authentication systems are presented in the subsequent sections. Block diagram of the ensemble classifier design is illustrated in Figure 1.

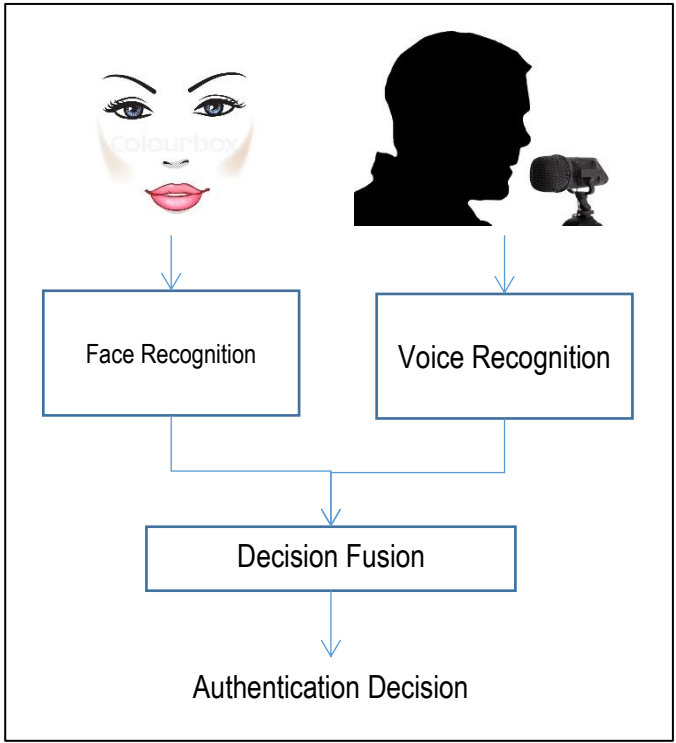

**Figure 1.** Bimodal authentication system block diagram

## II. Development of Face Recognition System

This section presents the steps in processing the raw face image data to extract features that form the inputs to a machine learning classifier, and the development of a classifier algorithm.

### A. *Face detection*

The first step in the face identification is the detection of a face or more specifically determining the coordinates of a face in the larger two-dimensional image. We employed the Viola-Jones algorithm to detect faces in a given image through its implementation in the OpenCV (Open Source Computer Vision) library [1]. The Viola-Jones object detection framework facilitates Haar-like features to be extracted from a face image as the initial processing step. Figure 2 shows an example of Haar-like features of a face image. Even though calculation of the Haar-like features is fast and efficient [11], a 24×24 pixel image window has 180,000 possible such features. In practice however, only a very small number of Haar-like features are needed. Given that there are potentially tens of thousands of Haar-like features for a face image, it is imperative, for computational efficiency purposes, to extract the "strong" ones that can be useful in detecting the face in an image. The Viola-Jones algorithm employs the AdaBoost machine learning algorithm to assign weights to all possible Haar-like features. Those features with highest weight values are considered as "strong" Haar-like features. Next, the *x* and *y* coordinates of the extracted Haar-like features are mapped back to the original image to extract the (*x*, *y*) coordinates of the detected face as laid out on a two-dimensional plane of face image. The mapping of coordinates as suggested results in a face rectangle from the forehead to the chin, and from the left to the right ear.

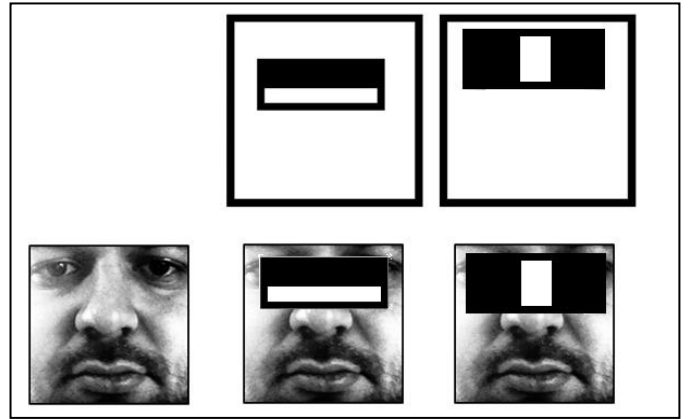

**Figure 2.** Haar-Like features superimposed on a face image

### B. *Preprocessing the face image*

The extracted face image will undergo several steps that entail processing the image to a state that a face image is ready to be used for training and recognition by a classification algorithm. These pre-processing steps, in order, include the following:
   a) eye detection,
   b) geometrical transformation and cropping,
   c) separate histogram equalization for left and right sides,
   d) smoothing, and
   e) application of an elliptical mask.

The positioning of eyes has to be nearly horizontal in a detected face image for the eye detection step to succeed. The regions where the eyes are most likely located can be determined using the following approach. A classifier that detects individual eye sub images runs independently on each eye region rectangle to locate the eyes [26]. The left and right eye region rectangles, named LERR and RERR respectively, are then defined where it is assumed that these two rectangle image regions have identical dimensions without loss of generality. Figure 3 shows the LERR and RERR of the left and right eye regions detected on a sample face image. After that, the center point for each detected eye is calculated (for use in the next step), where the "center" is the midpoint between the edges of each eye region as seen in Figure 3.

After detecting the eyes, the next step is to apply cropping on the image to remove face image "noise" (such as the hair, ears and part of the forehead) for improved recognition results. This processing step is performed using the Affine transformation [2]. To perform scaling and rotation of the image through Affine transformation, two matrices have to be formed. Once these two matrices, one for scaling and a second one for rotation, are generated, they are multiplied by the face image matrix resulting in a scaled and rotated image.

The rotation matrix requires calculation of values for a number of parameters, namely the Euclidean distance between the left and right eye, and the angle of the face rotated away from the *x*-axis (whose tangent is the quotient of the two distances) to determine the rotation needed (in degrees). The scaling ratio is calculated as follows. It is possible to generalize through hand measurements that the left eye center is typically at 14% to 19% of the detected face image. The standard face image height and width in this study is 70 units (pixels) as this value was found to be satisfactory for good performance by the set of detection and recognition algorithms without incurring excessive computational burden. Accordingly, the scaling ratio to scale down the input face image to 70 pixel units is calculated. After applying the Affine transformation, the result will be an image scaled down to 70×70 pixels and rotated to make the eyes horizontally aligned.

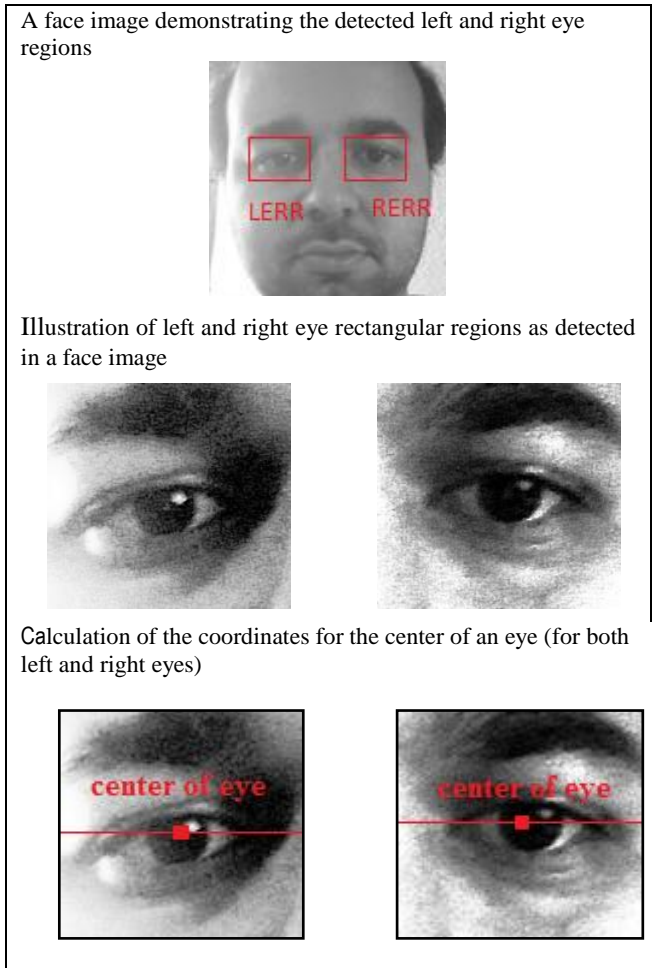

**Figure 3**. Starting with a face image the sequence of processing that culminates with the calculation of the "center of eye" for both left and right eye regions.

In real-world conditions, it is likely to have strong lighting on one-half of the face and weak lighting on the other. This might have a major adverse effect on the performance of a face recognition algorithm, as the left- and right-hand sides of the same face image will appear very different. Therefore, it is necessary to perform histogram equalization separately on the left and right halves of the face, to have standardized brightness and contrast on each side. Simply applying histogram equalization on the left half and then again on the right half of a face image would create a very distinct edge in the middle because the average brightness is likely to be different on the left and the right side. Therefore, to remove this likely boundary or edge, it is necessary to apply the two histogram equalizations gradually from the left and right-hand sides towards the center and combine it with an entire-face histogram equalization. This way the left-hand side will use the left histogram equalization, the right-hand side will use the right histogram equalization, and the center will use a smooth mix of left or right values and the whole-face equalized value [2]. To perform the separate histogram equalization for left and right sides of the face, we need copies of the whole face equalized as well as the left half equalized and the right half-equalized. This process standardizes the brightness and contrast on both the left and right hand side of the face independently. It helps significantly reduce the effect of different lighting on the left- and right-hand sides of the same face image.

In order to achieve a smoother face image, the bilateral filtering is applied for the next step. The main parameters for the bilateral filter are set as follows: $\sigma_d = 2$ pixels and $\sigma_r = 20$ gray levels [9]. We selected the value of 2 pixels because we need to smooth the pixel noise but not the large image regions [2]. As for the 20 gray levels, it is selected because the previous step, histogram equalization, would have increased the pixel noise, and a 20 gray level will be sufficient to counter the noise affect.

Most of the image background, forehead, and hair were removed when the geometrical transformation was applied. One can also apply an elliptical mask to remove some of the corner regions such as the neck, which might be in shadow due to the face, particularly if the face is not looking perfectly straight towards the camera [2]. Figure 4 shows an example image demonstrating the effect of the elliptical mask, and how it removes unwanted parts of the forehead and hair. To create the elliptical mask, one can draw a black-filled ellipse with the horizontal length of 70 pixels and a vertical radius of 120 pixels. The 70 and 120 pixels are chosen for the horizontal length and the vertical radius, respectively, based on the 70x70 pixel size of the scaled down face image.

*C. Face recognition algorithm*

The collection of face images for each subject is stored in templates using the Eigenfaces algorithm, which is a Principle Component Analysis (PCA) based technique [2]. Eigenfaces refers to an approach of appearance-based face recognition. It captures the variations in a collection of face images and uses this information to encode and compare images of individual faces in a holistic manner. In contrast with the other techniques which focus on particular features of the face, and therefore limiting the information being used, the Eigenfaces method is able to employ much more information through classifying the faces based on general facial patterns. Given that the entire face is analyzed, it is reasonable to conclude that the Eigenfaces approach is more effective than the feature-based approach because of its use of more information.

The main idea for the use of PCA for face recognition is to express the large one-dimensional vector of pixels constructed from two-dimensional facial images with the help of compact principal components of the feature space. This is called Eigenspace projection. Eigenspace projection is calculated by identifying the eigenvectors of the covariance matrix derived from a set of facial images or vectors. PCA is applicable to face recognition because face images are similar to each other. In the PCA approach, if the database consists of *M* images, each represented by an *N*×*N* matrix of real numbers, then the covariance matrix **C** of the images will yield $N^2$ eigenvectors and eigenvalues, as the covariance matrix **C** has dimensions of $N^2 \times N^2$ [7].

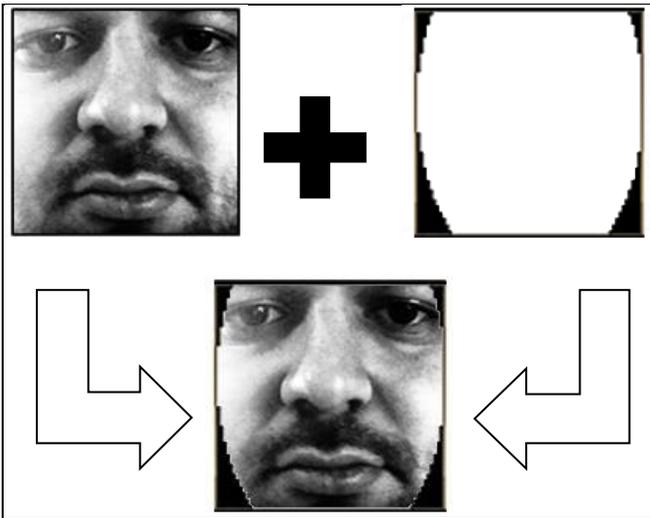

**Figure 4**. Illustration showing the effect of the elliptical mask and how it removes unwanted regions such as the forehead and hair

The application of the Eigenfaces algorithm for authentication requires valid users to provide a set of face images to create the so-called Eigenfaces and the computation of the average image separately for each user, and storage of the same in a secure database. When a user tries to authenticate, the user identification system will capture a still image of the user's face. That image will go through all the preprocessing steps previously described. Then, the face recognition algorithm will select among the registered subjects (users) the one with the average image that is the "most similar" to the face image of the subject who is currently being authenticated. The "similarity" is based on the "distance" between the two images, which is calculated using the Frobenius norm. If the distance between the most similar registered subject and the subject who is being authenticated is less than a threshold value then the user will be authenticated otherwise denied access. For an authenticated user, the distance value can be used as a confidence score for further processing.

## III. Development of Voice Recognition System

User authentication through voice identification is a feasible and practical option to enhance a secure access protocol since it is now commonplace to have high-fidelity microphones on various computing platforms including desktops, notebooks, tablets and smartphones. The broader perspective of integrating a voice recognition framework into a multi-modality user authentication system is poised to facilitate higher confidence levels in the recognition and identification processes. Voice identification entails four primary processing steps as follows:
  a) Signal acquisition,
  b) Feature extraction,
  c) Speaker modeling, and
  d) Speaker recognition.

*A. Signal acquisition*

The first step in voice identification is recording voice through microphones and using analog-to-digital converters to store the recordings in digital format. Voices may be recorded using a 16 KHz sampling rate, which is one of the typical frequencies [18] since the human ears respond to signals within the frequency spectrum covering the range from 20 Hz to 20 KHz. Recording may be performed on a single-channel (mono) to accommodate those devices with only one microphone. For individually spoken words, recording for a period of 3 seconds is sufficient to capture the spoken words completely as well as having sufficient data for the subsequent voice training and recognition [12-19].

*B. Feature extraction*

After acquiring the speech signal, the next step is to extract the needed features to generate the "voiceprint" that would uniquely identify a specific user. The speech signal is a slowly time-varying waveform, and when examined over a sufficiently short period of time (say in the range of 20 to 30 milliseconds), its characteristics are fairly stationary [12] [13]. However, over longer periods of time (0.5 second or more) the signal characteristics change to reflect the different speech sounds being made. There exist many possibilities of representing the speech signal parametrically [12]. Two such prominent methods are the Linear Prediction Coding (LPC), and the Mel-Frequency Cepstral Coefficients (MFCC).

The MFCC is the best known and most popular method for the extraction of voice features [12]. MFCCs are based on the known variation of the human ear's critical bandwidths with respect to frequency. Filters spaced linearly at low frequencies and logarithmically at high frequencies have been used to capture the phonetically important characteristics of speech. A "Mel" is a unit of measure based on the human ear's perceived frequency. This is expressed in the Mel-frequency scale; the spacing is linear at low frequencies (below 1000 Hz) and is logarithmic at high frequencies (above 1000 Hz) [20]. The Mel-frequency scale is appropriate for speech because human ear perceives sounds in a nonlinear fashion, allowing the MFCC features to be extracted similar to how human ears hear speech. MFCCs are shown to be robust in the presence of the variation of the speaker's voice and noise in the surrounding environment. The multistep procedure to derive MFCCs for a speech signal is as follows:
  a) Divide the signal into short frames.
  b) Compute the Fourier transform of (a windowed excerpt of) a signal.
  c) Map the powers of the spectrum obtained above onto the Mel scale, using triangular overlapping windows.
  d) Compute the logs of the powers at each of the Mel frequencies.
  e) Compute the discrete cosine transform of the list of Mel log powers, as if it were a signal. The MFCCs are the amplitudes of the resulting spectrum.

First step is frame blocking. The continuous speech signal is blocked into $R$ frames of $N$ samples of approximately 30 milliseconds. To be able to extract as many features as possible from a speech sample, the overlapping of frames technique may be used [14]. With an overlap of 50%, the first frame consists of the first $N$ samples. The second frame begins $M$ samples after the first frame, and overlaps the first frame by $N – M$ samples. Similarly, the third frame begins $2M$ samples after the first frame (or $M$ samples after the second frame) and overlaps the second frame by $N – 2M$ samples. This process continues until all the speech is accounted for [15]. Determining the number of samples for frame blocking, both the time and the frequency domains have to be considered. For a 16 KHz sampling rate and a 30-millisecond frame, it is reasonable to consider *N*=512 samples. That will result in a good tradeoff between the number of frames to process and amount of data at various frequencies.

Frame blocking results in the signal to be distorted at the start and end of a frame. To minimize this distortion, windowing will need to be performed in the subsequent step. Two windowing options commonly used during the frequency analysis of speech sounds are Hamming and Hanning windows [16]. These windows are formed by inverting and shifting a single cycle of a cosine function so as to constrain the values to a specific range: namely [0, 1] for the Hanning window; [0.054, 1] for the Hamming window. Based on the same function template shown below, the Hamming window employs $\varphi = 0.54$ while the Hanning window uses $\varphi = 0.5$:

$$w(n) = \varphi - (1-\varphi) \times \cos\left(\frac{2\pi n}{N-1}\right), \quad (1)$$

where *N* represents the width, in samples, of a discrete-time, symmetrical window function $w(n)$, $0 \le n \le N-1$. For this project, the choice is somewhat arbitrary due to comparable performances of these two options and hence, the most popular method was chosen, which is the Hamming window [16].

The approximation for "mel" from frequency can be expressed as

$$mel(f) = 2595 \times \log_{10}(1 + \frac{f}{700}) \quad (2)$$

where *f* denotes the real frequency, and *mel*(*f*) denotes the perceived frequency. The Mel-frequency warping is realized through filter banks. Filter banks are usually implemented in the frequency domain (instead of the time domain). The center frequencies of the filters are evenly spaced on the frequency axis. However, in order to mimic the human ears' perception, the warped axis according to the Mel non-linear function $mel(f)$, is implemented. The most commonly-used filter shape is triangular. The output of the *i*[th] filter $Y(i)$ is calculated using

$$Y(i) = \sum_{j=1}^{N} S(j)\psi_i(j), \quad (3)$$

where $S(j)$ is an *N*-point magnitude spectrum, and $\psi_i(j)$ is the sampled magnitude response of an *M*-channel filter bank $(i = 1, 2, ..., M)$. The output of the *i*[th] filter can be observed as the magnitude response of speech signal in that frequency region weighted by the filter response. The last step before getting the Mel-Frequency Cepstral Coefficients is the Inverse Discrete Fourier Transform (IDFT). Normally, the Discrete Cosine Transform (DCT) is performed instead of IDFT, since $\log Y(i)$ is symmetrical about the Nyquist frequency. Therefore, MFCCs are calculated as:

$$c_s(n,m) = \sum_{i=1}^{M} (\log Y(i)) \times \cos\left(i\left(\frac{2\pi}{N'}\right)n\right), \quad (4)$$

where $N'$ is the number of points used to compute the IDFT.

*C. Speaker modeling*

The MFCC feature vectors are typically scattered all across the associated space. To make sense out of them, we need to create a model that is implemented through a template that identifies the speaker uniquely based on the MFCC vectors. We use vector quantization to create the template for a particular user, as it would be impractical to store every single feature vector which we generate from the training utterance [12]. Vector quantization (VQ) is a well-known technique for signal processing which allows the modeling of density functions by the distribution of vectors. It works by dividing a large set of points (vectors) into groups having approximately the same number of points closest to them. Each group is represented by its centroid point as in the K-means clustering algorithm.

The specific VQ technique employed for this project is called the Linde–Buzo–Gray (LBG) algorithm. LBG algorithm is similar to the *K*-means clustering algorithm which takes a set of vectors $S = \{\mathbf{x}_i \in R^d | i = 1, 2, ..., n\}$ as input and generates a representative and much smaller subset of vectors $C = \{\mathbf{c}_j \in R^d | j = 1, 2, ..., K\}$ or codewords, where $R^d$ is the input set and $K$ is the desired number of clusters with $K<<n$. Initialization is an important step in the codebook estimation. The approach used in this project for the initial codebooks is taking the average of all of the training vectors and creating a vector with single element (holding the average). The selection of the $K$ value is important as a small number can result in a very non-descriptive codebook of the speaker, and a very big number could cause in an overly quantized codebook of the speaker's MFCCs. Through exploratory simulation-based study results and considering the requirements of this project, we determined that setting $K = 8$ is a good choice.

*D. Speaker recognition*

After the first three steps (namely signal acquisition, feature extraction and speaker modeling) are completed, the user's "voiceprint" is captured in the form of a codebook. The codebook will be stored in the user's profile record completing the system development phase.

Following the deployment of the system, when a user is in the process of authenticating, his or her voice sample is captured. Afterwards, the speech signal is subjected to feature extraction and speaker modeling processes generating the codebook of the authenticating user. The recognition system will then query the user profiles, comparing the codebook of the authenticating user against the previously-stored codebooks belonging to known users. The comparison between the codebooks in the database and the codebook of the authenticating user is performed using the Euclidean distance. That user whose codebook has the smallest Euclidean distance to the authenticating user will be identified. The authenticating user will be granted access as that user if the Euclidean distance between these two codebooks is less than a preset threshold value. If the user is authenticated, the Euclidean distance value can be used as a (confidence) score for further processing.

## IV. Data Fusion for Bi-modal Biometric Authentication

The proposed ensemble design has two base classifier modules. Outputs of these two different modality classifiers need to be combined to generate a final authentication decision by the ensemble classifier. For biometric authentication designs that utilize multiple and different modality biometric classifiers, the transformation-based fusion technique is ideal as it makes it relatively easy to combine classifier outputs. Consequently, this study employs the transformation-based fusion technique to combine the outputs of two base classifiers. This fusion technique is only dependent on the score generated from each biometric (processing or classification) module. It combines the scores using the sigmoid function (used for score normalization), and thereby generating a final score that is robust (i.e. the method is not sensitive to outliers in the data) and computationally efficient.

The first step of processing is the normalization among the scores originating from different biometric classifiers. The process of score normalization consists of changing the scale parameters of the underlying match score distributions to ensure compatibility between multiple score types. To achieve normalization of scores, the double sigmoid function is one option since it results in a highly efficient and robust mapping as it is not sensitive to outliers in the data [3][10]. The normalized score is calculated through the following procedure. Let $\Phi_{j,k}$ denote the normalized score, $s_{j,k}$ denote the $k$-th match score output by the $j$-th biometric module with $k=1,2,\ldots,P$ and $j=1,2,\ldots,R$, where $P$ is the number of match scores available in the training set and $R$ is the number of biometric modules. Then, the formula for $\Phi_{j,k}$ is given as

$$\Phi_{j,k} = \begin{cases} \dfrac{1}{1+e^{-2\left(\frac{s_{j,k}-\tau}{\alpha_1}\right)}}, & \text{if } s_{j,k} < \tau \\[2ex] \dfrac{1}{1+e^{-2\left(\frac{s_{j,k}-\tau}{\alpha_2}\right)}}, & \text{otherwise} \end{cases}$$

(5)

where $\tau$ is the reference operating point; $\alpha_1$ and $\alpha_2$ denote the left and right boundaries of the region in which the function is linear. The double sigmoid function exhibits linear characteristics in the interval ($\tau-\alpha_1$, $\tau-\alpha_2$). While the double sigmoid normalization maps the scores into the [0, 1] interval, it requires careful tuning of the parameters $\tau$, $\alpha_1$ and $\alpha_2$ to obtain good efficiency. Generally, $\tau$ is chosen to be some value falling in the region of overlap between the genuine and impostor score distributions, and $\alpha_1$ and $\alpha_2$ are set so that they relate to the extent of overlap between the two distributions toward the left and right of $\tau$, respectively.

Once the match scores output by multiple biometric modules are normalized, they can be combined using a fusion operator. One widely used example is the weighted sum of scores (WSS) which combines two or more normalized scores with different weights into a single one:

$$WSS_k = \sum_{i=1}^{M_k} \omega_i \Phi_{i,k},$$

(6)

where $M_k$ is the number of scores for the $k$-th match, and $\omega_i$ is the weight of $i^{th}$ score.

During user authentication, both the face and the voice biometric modules (classifiers) implement the following generic steps:

1. Capture, respectively, the face image and voice print of an individual;
2. Extract feature sets;
3. Compare those features against the same user's face image-based and voice print based templates which are previously-stored in a database; and
4. Generate a decision regarding the identity of the user.

Both classifiers also generate a separate "score" representing a certain degree of confidence in the classification decision rendered. These two scores can readily be utilized by a transformation based fusion technique to formulate an authentication decision, which is the approach adopted in this study.

## V. Simulation Study

Simulation study first presents the performance assessment and evaluation of two base classifiers individually. This is followed by the comparative performance evaluation of the ensemble classifier.

### A. Face identification module

Performance evaluation of the face identification module was accomplished through two benchmark data sets [3,4] A data set must have substantial variation with respect to a number of attributes like lighting conditions for the face image capture, subject ethnicity, gender, age, types of poses, and test subject count: this is so that the real-life scenarios can be mimicked. For this purpose, two still-image face databases were identified, which are the Yale Extended Face database [4] and the NIST FERET database [3], as having desirable set of attributes. Tables 1 and 2 present the characteristics of the Yale Extended Face and the NIST FERET databases, respectively.

| Data Set Property | Count |
|---|---|
| Number of Subjects | 28 |
| Male Subjects | 20 |
| Female Subjects | 8 |
| Ethnicity Distribution | Caucasian, African-American, Latin-America, Asian |
| Number of Poses Per Subject | 9 |
| Number of Lighting Conditions Per Pose | 65 |
| Total Number of Images | 16380 |

*Table 1.* Yale Extended Face Image Database Attributes

In both databases, some face samples are not valid: for example, only half of the face is visible or lighting is very low that the image is too dark for a large part. The face identification system requires detection of both eyes to confirm the validity of a face image. Requirements for a face image to be considered valid for training and authentication are as follows (this is enforced by the fact that any detected face, for it to be valid, both eyes must be located along a line):

- Face has to be looking straight toward the camera, with both eyes visible.
- Lighting must be such that the face can be identified in the image: face image cannot be so dark that the face is completely black nor too bright that the face is completely blurred with no features visible. Very dark lighting is defined by the fact that no edges can be seen/detected in an image due to darkness, causing the face to blur in with the background. Therefore, no face or eyes can be detected.

| Data Set Property | Count |
|---|---|
| Number of Subjects | 994 |
| Male Subjects | 401 |
| Female Subjects | 324 |
| Ethnicity – White | 426 |
| Ethnicity – African American | 57 |
| Ethnicity – Middle Eastern | 41 |
| Ethnicity – Asian | 141 |
| Ethnicity – Pacific Islander | 9 |
| Ethnicity – Hispanic | 47 |
| Ethnicity – Native American | 2 |
| Age 20-30 years old | 17 |
| Age 31-50 years old | 547 |
| Age 51-70 years old | 135 |
| Age 70-90 years old | 26 |
| Number of Poses Per Subject | Varies |
| Number of Lighting Conditions Per Pose | Varies |
| Total Number of Images | 12328 |

*Table 2*. NIST FERET Face Image Database Attributes

Accordingly, this "face image validity" feature of the design requires the exclusion of invalid face images from the two databases. The following criteria are used to exclude face images in the two databases:
- Face images with only the side profiles visible are not considered.
- Faces with extreme lighting conditions (such as completely dark) are removed.
- Images with faces looking away from the camera, so that both eyes are not clearly visible in the image are excluded.
- Subjects that have fewer than 10 acceptable face image samples are not considered for inclusion in the study as this number is considered a good estimate for learnability by machine learning algorithms for classification.

Face identification system parameters and their values used in the simulation study are as presented in Table 3. The "Image Scaling Width" parameter is used to scale down (or up) the images to 320 pixels in width. This value is established so that large images will not cause the face detector to spend a long time searching for the face. The 320-pixel value was chosen empirically through testing performance of the face detector as reported in [2]. The "Detected Face Width" and "Detected Face Height" parameters are also used to scale down (or up) the face image. Values for these two parameters are set so that large size face images (with high pixel counts) will not cause the face recognizer to spend excessive amount of time in training and recognition [2]. The "Eigenfaces Distance Threshold" parameter facilitates the recognizer to consider a recognized image as a false or true result. The value for this parameter was chosen based on exploratory simulation work performed on the datasets referred to herein, and 2800 was determined to be the optimal value. A very large value for this parameter will result in the false positive rate to increase. On the other hand, a very low value will result in the false negative rate to increase. Figure 5 shows a graph of performance vs. the threshold value for this parameter. The graph was generated using the Yale Extended Face Database. Values for the "Face Samples per Subject for Training" and "Face Samples per Subject for Testing" parameters were determined to mimic the normal usage of the system following its deployment in the field.

| Parameters | Values |
|---|---|
| Image Scaling Width | 320 pixels |
| Detected Face Width | 70 pixels |
| Detected Face Height | 70 pixels |
| Eigenfaces Distance Threshold | 2800 |
| Face Samples per Subject for Training | 20 to 30 |
| Face Samples per Subject for Testing | 10 |

*Table 3*. Face Identification Module Parameters and Values

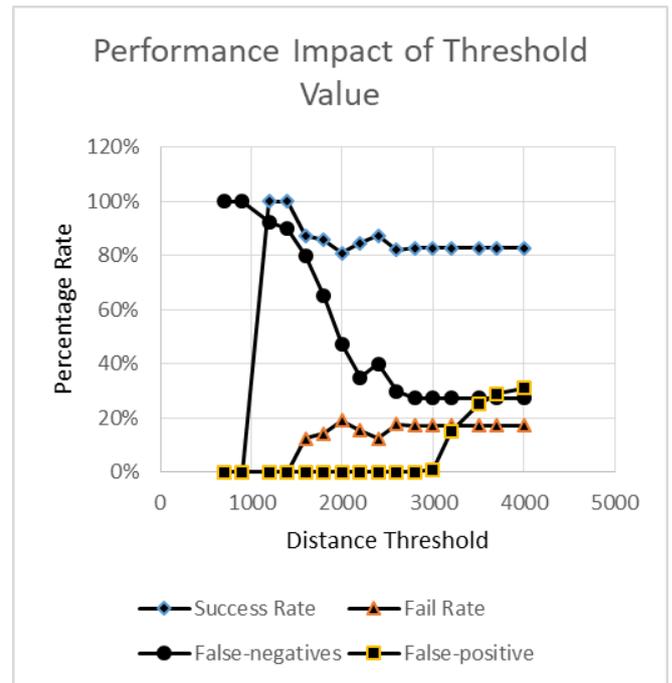

**Figure 5**. Impact of threshold value on performance

Performance of the authentication algorithm is characterized through the confusion matrix as well as the Accuracy (AC), True Positive Rate (TP), False Positive Rate (FP), True Negative Rate (TN), False Negative Rate (FN), and Precision (P) metrics. Tables 4 and 6 present the confusion matrices for the Yale Extended and NIST FERET databases, respectively, while Tables 5 and 7 present the values for the six performance metrics for each of the same two databases.

From Tables 5 and 7, the true positive rate is at 91.2% for the Yale and 98.4% for the FERET databases suggesting that the system can authenticate the legitimate user at a reasonably high rate. True negative rates are 87.1% and 97% for these databases suggesting that the system denies access correctly to unauthorized users. On the other hand, false positive rates are 12.8% and 2.9% where the system incorrectly authenticates unauthorized users. False negative rates are 8.7% and 1.5% for these databases, where the system incorrectly denies authentication to authorized users. In comparison, Slavković, et al. in [8] and Turk et al. in [9] report the true positive rates of 92.5% and 96% for the two databases, namely Yale Extended and NIST FERET, respectively. Therefore, although performance of the system implemented in this study is competitive with those reported in the literature, there is still room for improvement. There are important differences for the face images belonging to these two databases, namely the Yale Extended database and the NIST FERET database, to explain in part the performance differences observed. The NIST FERET focuses more on the position of the face toward the camera, while the Yale Extended database focuses more on the lighting conditions. Given these differing emphases, the observed variation for performance is expected.

|  |  | Recognized as | |
|---|---|---|---|
|  |  | NRU | RU |
| **Actual** | NRU | 15321 | 2256 |
|  | RU | 1352 | 14124 |

Table 4. Confusion matrix for the Yale Extended data (NRU: Non-Registered User, RU: Registered User)

| Performance Indicator | Value |
|---|---|
| Accuracy | 89.08% |
| True Positive Rate | 91.26% |
| False Positive Rate | 12.84% |
| True Negative Rate | 87.17% |
| False Negative Rate | 8.74% |
| Precision | 86.23% |

Table 5. Performance of face identification system on the Yale Extended data

|  |  | Recognized as | |
|---|---|---|---|
|  |  | NRU | RU |
| **Actual** | NRU | 12122 | 370 |
|  | RU | 192 | 11958 |

Table 6. Confusion Matrix for the NIST FERET data (NRU: Non-Registered User, RU: Registered User)

| Performance Indicator | Value |
|---|---|
| Accuracy | 97.72% |
| True Positive Rate | 98.42% |
| False Positive Rate | 2.96% |
| True Negative Rate | 97.04% |
| False Negative Rate | 1.58% |
| Precision | 96.99% |

Table 7. Performance of the face identification system on the NIST FERET data

*B. Voice identification module*

We used a benchmark dataset, the English Language Speech Database for Speaker Recognition (ELSDSR) [17] that possesses a high degree of variation – with respect to gender, ethnicity and age and others – and therefore can mimic real-life scenarios to evaluate the performance of voice identification module. Table 8 highlights the main characteristics of this database.

For the development of the voice identification module, parameters and their values used in the simulation study are presented in Table 9. The "Recording sample rate" parameter indicates the signal sampling value for the voice recording, which is 16 KHz. This value is set to cover the frequency components of the human ear hearing range [18]. The "VQ LBG Cluster size" parameter is used to configure the LBG algorithm to determine how many code words are in the codebook. A value of 8 is chosen for this parameter based on exploratory testing results (along with using 16 KHz sampling rate per second and using 512 samples per frame). Results show that having fewer than 8 clusters will cause the codebook generated by the LBG algorithm not to have enough features and more than 8 clusters will result in weak feature values. Having the MFCC vectors divided into fewer than 8 clusters will have many vectors average out and lose their details, while more than 8 clusters will have the MFCC vectors divided into more clusters, causing code words to be generated with fewer number of vectors, resulting in less accurate averages. The "Number of samples per frame" parameter is used to divide the recorded samples into frames, and for this study, its value is set as 512 samples. Values for "Voice Samples per Subject for Training" and "Voice Samples per Subject for Testing" parameters were chosen to mimic the typical usage of the system following its deployment in the field. Each user registered with the voice identification system using 5 to 7 randomly selected voice recordings of that specific user. The remaining voice recordings of the same user, which were not used during registration, were reserved for the testing phase.

| Data Set Property | Value |
|---|---|
| Number of Subjects | 22 |
| Voice samples per subject | 9 |
| Male Subjects | 12 |
| Female Subjects | 10 |
| Age range | 24 to 63 |
| Ethnicity | Varies [31] |
| Total Number of voice samples | 198 |

Table 8. ELSDSR database features

| Simulation Study Parameters | Value |
|---|---|
| Recording sampling rate | 16000 samples per second (Hz) |
| Recording channel | 1 (monaural, single-channel with 1 microphone) |
| VQ LBG Cluster size ($K$) | 8 |
| Number of samples per frame | 512 (Hz) |
| Voice samples per subject for training | 5 to 7 |
| Voice samples per subject for testing | 2 |

Table 9. Simulation study parameters for voice identification module

For the ELSDSR database, the confusion matrix and the performance metrics of Accuracy (AC), True Positive Rate (TP), False Positive Rate (FP), True Negative Rate (TN), False Negative Rate (FN), and Precision (P) assume the values presented in Tables 10 and 11. Table 11 shows that the true positive rate is at 98.98% indicating that the system can authenticate the correct user at a high rate. The true negative rate is 98.36% suggesting that the system denies access correctly to unauthorized users. Given the false positive rate value of 1.63%, the system incorrectly authenticates less than 2 unauthorized users per 100 users. A false negative rate of 1.01% results in denial of authentication of one user out of 100 users. Jiahong, et al. in [19] reported the true positive rate of 98.0% on the ELSDSR dataset for their study. This indicates that the voice identification design in this study is competitive with those reported in the literature.

|  |  | **Recognized as** | |
|---|---|---|---|
|  |  | NRU | RU |
| **Actual** | NRU | 181 | 3 |
|  | RU | 2 | 196 |

Table 10. Confusion matrix for the ELSDSR data (NRU: Non-Registered User, RU: Registered User)

| Performance Indicator | Value |
|---|---|
| Accuracy | 98.69% |
| True Positive Rate | 98.98% |
| False Positive Rate | 1.63% |
| True Negative Rate | 98.36% |
| False Negative Rate | 1.01% |
| Precision | 98.49% |

Table 11. Performance of voice identification system on the ELSDSR data

*C. Bi-modal authentication system: ensemble classification*

A distributed client-server software system was created to facilitate the simulation study. Testing entailed using smartphones, multiple users, and authentication over actual GSM networks, and real-time decision-making performance measurements. Figure 6 illustrates the implementation diagram.

Two base classifiers developed for face and voice recognition in the previous sections were incorporated into an ensemble classification framework where the combiner was based on a transformation-based score fusion algorithm. The data for performance evaluation included the Yale Extended [24] and NIST FERET databases [23] for the face images and the ELSDSR database [25] for the voice recordings. Subjects with voice recording in the ELSDSR dataset were associated with a face subject from Yale Extended or NIST databases at random. Since there are more subjects in the Yale and NIST databases (1022 subjects) than the ELSDSR database (22 subjects) performance evaluation was repeated 100 times; for each iteration, the 22 voice subjects in the ELSDSR database are associated with different face subjects in either Yale or NIST face databases at random.

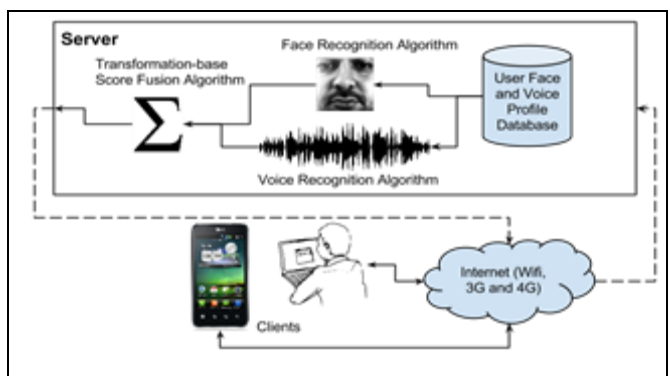

**Figure 6.** Diagram illustrating the server/client architecture implementation

Values of parameters in Equations 5 and 6 as they were employed in the simulation study are presented in Table 12. The threshold parameters ($\tau$) are used in the double sigmoid score normalization function for the face and voice recognition base classifier outputs, which is defined by Equation 5. The face and voice recognition left and right edge parameters, $\alpha_1$ and $\alpha_2$, are the minimum and maximum values of each face and voice recognition pattern distances as calculated by the corresponding base classifier modules. Values of parameters appearing in Equation 5 are determined through an empirical trial and error process. The parameter $\omega$ is used to assign weight values to the face and voice recognition scores during the "weighted sum of scores" (WSS) based calculation as in Equation 6. The weight value $\omega$ for the face recognition classifier input to the fusion module is smaller than that of the voice recognition module (35% vs. 65%, respectively) because face recognition is affected more by the surrounding environmental noise (such as lighting conditions) compared to the voice recognition. Incidentally, the voice recognition module performs feature extraction using the MFCCs, which is known to be less susceptible to the variation of the speaker's voice and the noise in the surrounding environment. Figures 7 and 8 illustrate the registration and authentication process flows on the client and server sides, respectively.

| Parameter | Face | Voice |
|---|---|---|
| Score threshold ($\tau$) | 2800 | 2.6 |
| Left boundary ($\alpha_1$) | 200 | 0.3 |
| Right boundary ($\alpha_2$) | 3400 | 3.1 |
| Score weight ($\omega$) | 0.35 | 0.65 |

Table 12. Parameter values for score normalization and fusion.

The performance assessment and evaluation was conducted by generating a score from each of the face and voice recognition base classifiers, and then fusing the scores into a single score to generate an authentication decision. The confusion matrix and the performance metrics of accuracy, true positive rate, false positive rate, true negative rate, false negative rate, and precision assume the values presented in Tables 13 and 14.

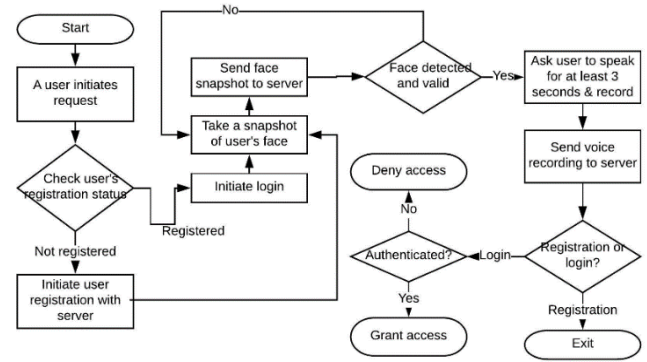

**Figure 7**. Diagram illustrating the registration and authentication process flows on the client side

The true positive rate for the ensemble classifier is at 99.22%, indicating that the system can authenticate the legitimate user at the rate of 99 out of 100 cases. The true negative rate is 99.28% for both databases showing that the system denies access correctly to unauthorized users for 99 out of 100 cases. Consequently, the false positive rate shows 0.71% where the system incorrectly authenticates less than one out of every 100 unauthorized users. The false negative rate of 0.84% suggests that the system incorrectly denies authentication to less than one in every 100 authorized users. In comparison with the performances of each biometric module deciding on its own, fusion score based ensemble classification improved the performance significantly overall. Four performance indicators, namely accuracy, true positive rate, true negative rate and precision are all at 99%, while false positive and negative rates are less than 1%. The overall performance of the ensemble classifier as a bimodal biometric authentication system is promising.

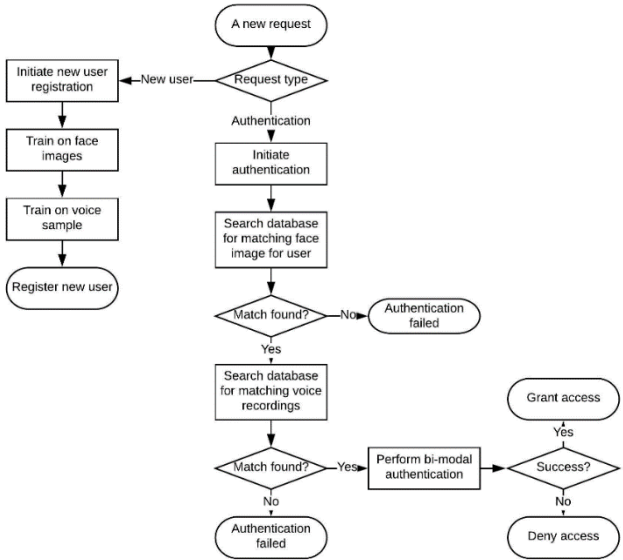

**Figure 8.** Diagram illustrating the registration and authentication process flows on the server side

|  |  | Authenticated As | |
|---|---|---|---|
|  |  | Unknown User | Known User |
| **Actual** | Unknown User | 27443 | 198 |
|  | Known User | 222 | 26082 |

*Table 13.* Confusion matrix for ensemble classifier

| Performance Indicator | Ensemble Classifier | Face Classifier | Voice Classifier |
|---|---|---|---|
| Accuracy | 99.22% | 89.08% (Yale), 97.72% (NIST) | 98.70% |
| True Positive Rate | 99.15% | 91.26% (Yale), 98.42% (NIST) | 98.98% |
| False Positive Rate | 0.71% | 12.84% (Yale), 2.96% (NIST) | 1.63% |
| True Negative Rate | 99.28% | 87.17% (Yale), 97.04% (NIST) | 98.36% |
| False Negative Rate | 0.84% | 8.74% (Yale), 1.58% (NIST) | 1.01% |
| Precision | 99.24% | 86.23% (Yale), 96.99% (NIST) | 98.50% |

*Table 14.* Performance of the ensemble classifier

## VI. Conclusions

This study presented design and performance evaluation of a bi-modal biometric user authentication system based on a machine learning ensemble classifier. Two base classifiers, one for face and a second one for voice identification, are employed by the ensemble classifier. Design and performance evaluation of individual base classifiers for face and voice recognition is presented. This is followed by the ensemble classifier design and performance evaluation. Three benchmark datasets, namely NIST FERET, and Yale Extended for face images, and ELSDSR for voice, are used in the study. Simulation results indicate that ensemble classifier improves upon performances of individual classifiers and performs at a promising level for consideration towards deployment in a real life context.

## Author Biographies

**Firas Abbaas** received both his BS & MS in Computer Science and Engineering from the University of Toledo in 2011 and 2014, respectively. He has been working as a software developer for the management consulting firms since graduation. He is interested in implementation of a variety of state of the art software engineering technologies including Artificial Intelligence.

**Gursel Serpen** received his PhD in Electrical Engineering with specialization in Computer Engineering from the Old Dominion University in 1992. His research interests are in the areas of AI, and Machine Learning with particular emphasis in Artificial Neural Networks, and their applications mainly in Big Data, secure computing, adaptation and optimization, bio-medical informatics, simulation and robotic path planning.